\documentstyle[12pt]{article}

\input epsf
\ifx\epsffile\undefined
\newlength{\epsfysize}
\def\epsffile#1#2#3#4]#5{}
\else
\fi

\catcode`\@=11
\long\def\@makefntext#1{
\protect\noindent \hbox to 3.2pt {\hskip-.9pt
$^{{\ninerm\@thefnmark}}$\hfil}#1\hfill}		%CAN BE USED

 \def\@makefnmark{\hbox to 0pt{$^{\@thefnmark}$\hss}}  %ORIGINAL

\def\ps@myheadings{\let\@mkboth\@gobbletwo
\def\@oddhead{\hbox{}
\rightmark\hfil\ninerm\thepage}
\def\@oddfoot{}\def\@evenhead{\ninerm\thepage\hfil
\leftmark\hbox{}}\def\@evenfoot{}
\def\sectionmark##1{}\def\subsectionmark##1{}}

\newcounter{sectionc}\newcounter{subsectionc}\newcounter{subsubsectionc}
\renewcommand{\section}[1] {\vspace{0.6cm}\addtocounter{sectionc}{1}
\setcounter{subsectionc}{0}\setcounter{subsubsectionc}{0}\noindent
	{\bf\thesectionc. #1}\par\vspace{0.4cm}}
\renewcommand{\subsection}[1] {\vspace{0.6cm}\addtocounter{subsectionc}{1}
	\setcounter{subsubsectionc}{0}\noindent
	{\it\thesectionc.\thesubsectionc. #1}\par\vspace{0.4cm}}
\renewcommand{\subsubsection}[1]
{\vspace{0.6cm}\addtocounter{subsubsectionc}{1}
	\noindent {\rm\thesectionc.\thesubsectionc.\thesubsubsectionc.
	#1}\par\vspace{0.4cm}}

\newcounter{appendixc}
\newcounter{subappendixc}[appendixc]
\newcounter{subsubappendixc}[subappendixc]

\renewcommand{\appendix}[1] {\vspace{0.6cm}
        \refstepcounter{appendixc}
        \setcounter{figure}{0}
        \setcounter{table}{0}
        \setcounter{equation}{0}
        \renewcommand{\thefigure}{\Alph{appendixc}.\arabic{figure}}
        \renewcommand{\thetable}{\Alph{appendixc}.\arabic{table}}
        \renewcommand{\theappendixc}{\Alph{appendixc}}
        \renewcommand{\theequation}{\Alph{appendixc}.\arabic{equation}}
        \noindent{\bf Appendix \theappendixc #1}\par\vspace{0.4cm}}

\def\abstracts#1{{
	\centering{\begin{minipage}{30pc}\tenrm\baselineskip=12pt\noindent
	\centerline{\tenrm ABSTRACT}\vspace{0.3cm}
	\parindent=0pt #1
	\end{minipage}}\par}}

\renewenvironment{thebibliography}[1]
	{\begin{list}{\arabic{enumi}.}
	{\usecounter{enumi}\setlength{\parsep}{0pt}
\setlength{\leftmargin 1.25cm}{\rightmargin 0pt}
	 \setlength{\itemsep}{0pt} \settowidth
	{\labelwidth}{#1.}\sloppy}}{\end{list}}

\newcounter{itemlistc}
\newcounter{romanlistc}
\newcounter{alphlistc}
\newcounter{arabiclistc}

\newcommand{\fcaption}[1]{
        \refstepcounter{figure}
        \setbox\@tempboxa = \hbox{\tenrm Fig.~\thefigure. #1}
        \ifdim \wd\@tempboxa > 5.5in
           {\begin{center}
        \parbox{5.5in}{\footnotesize\baselineskip=12pt Fig.~\thefigure. #1}
            \end{center}}
        \else
             {\begin{center}
             {\tenrm Fig.~\thefigure. #1}
              \end{center}}
        \fi}

\newcommand{\tcaption}[1]{
        \refstepcounter{table}
        \setbox\@tempboxa = \hbox{\tenrm Table~\thetable. #1}
        \ifdim \wd\@tempboxa > 6in
           {\begin{center}
        \parbox{6in}{\tenrm\baselineskip=12pt Table~\thetable. #1}
            \end{center}}
        \else
             {\begin{center}
             {\tenrm Table~\thetable. #1}
              \end{center}}
        \fi}

\def\@citex[#1]#2{\if@filesw\immediate\write\@auxout
	{\string\citation{#2}}\fi
\def\@citea{}\@cite{\@for\@citeb:=#2\do
	{\@citea\def\@citea{,}\@ifundefined
	{b@\@citeb}{{\bf ?}\@warning
	{Citation `\@citeb' on page \thepage \space undefined}}
	{\csname b@\@citeb\endcsname}}}{#1}}

\newif\if@cghi
\def\cite{\@cghitrue\@ifnextchar [{\@tempswatrue
	\@citex}{\@tempswafalse\@citex[]}}
\def\citelow{\@cghifalse\@ifnextchar [{\@tempswatrue
	\@citex}{\@tempswafalse\@citex[]}}
\def\@cite#1#2{{$\null^{#1}$\if@tempswa\typeout
	{IJCGA warning: optional citation argument
	ignored: `#2'} \fi}}

\def\fnt#1#2{\footnotetext{\kern-.3em
	{$^{\mbox{\sevenrm #1}}$}{#2}}}

 1
 1
 1

\font\tenbf=cmbx10
\font\tenrm=cmr10
\font\tenit=cmti10

\font\ninerm=cmr9

\begin{document}

\centerline{\tenbf COMPLETE WEAK-SCALE THRESHOLD CORRECTIONS}
\baselineskip=16pt
\centerline{\tenbf IN THE MINIMAL SUPERSYMMETRIC STANDARD MODEL}
\vspace{0.8cm}
\centerline{\tenrm J.~BAGGER, K.~MATCHEV AND D.~PIERCE}
\baselineskip=13pt
\centerline{\tenit Department of Physics and Astronomy}
\baselineskip=12pt
\centerline{\tenit The Johns Hopkins University}
\centerline{\tenit Baltimore, MD  \ 21218,\ \ USA}
\vspace{0.9cm}
\abstracts{
We compute the full set of weak-scale gauge and Yukawa
threshold corrections in the minimal supersymmetric standard model,
and use them to study the effects of the supersymmetric particle
spectrum on gauge and Yukawa coupling unification.}
\vfil

\def\o{\over}
\def\dr{\mbox{$\overline{\it DR}$}~}
\def\ms{\mbox{$\overline{\it MS}$}~}
\def\roughly#1{\raise.3ex\hbox{$#1$\kern-.75em\lower1ex\hbox{$\sim$}}}

\rm\baselineskip=14pt
\section{Introduction}

The idea of supersymmetric unification has received a significant
boost from the recent observation that the gauge couplings unify
in the minimal supersymmetric standard model.\cite{unified}
Indeed, this observation has given great impetus to the search
for supersymmetric particles at present and future accelerators.
In this talk we will assume that supersymmetry has been found,
and ask what the supersymmetric particle spectrum can tell us
about unification-scale physics.

To lowest order, the answer is not much, because the one-loop
renormalization group equations (RGE's) do not depend on the
physics of the unification scale.  To second order, the story is different.
This is because the two-loop RGE's must be used in conjunction with
the one-loop threshold corrections.  The threshold conditions depend
on the weak-scale supersymmetric spectrum {\it and} on the unknown
unification-scale physics.  Therefore to second order, the superparticle
spectrum is correlated with the physics of the unification scale.

In this talk we will report on results of a complete next-to-leading-order
analysis of supersymmetric unification.  Our results
are based on the full set of two-loop supersymmetric
renormalization group equations, together with the complete one-loop
threshold conditions, including finite parts.
We use a two-sided differential equation solver which allows us to
impose boundary conditions at the weak scale, $M_Z$, and at the
unification scale, $M_{\rm GUT}$.

\begin{figure}[t]
\epsfysize=2.5in
\hspace*{0.5in}
\epsffile{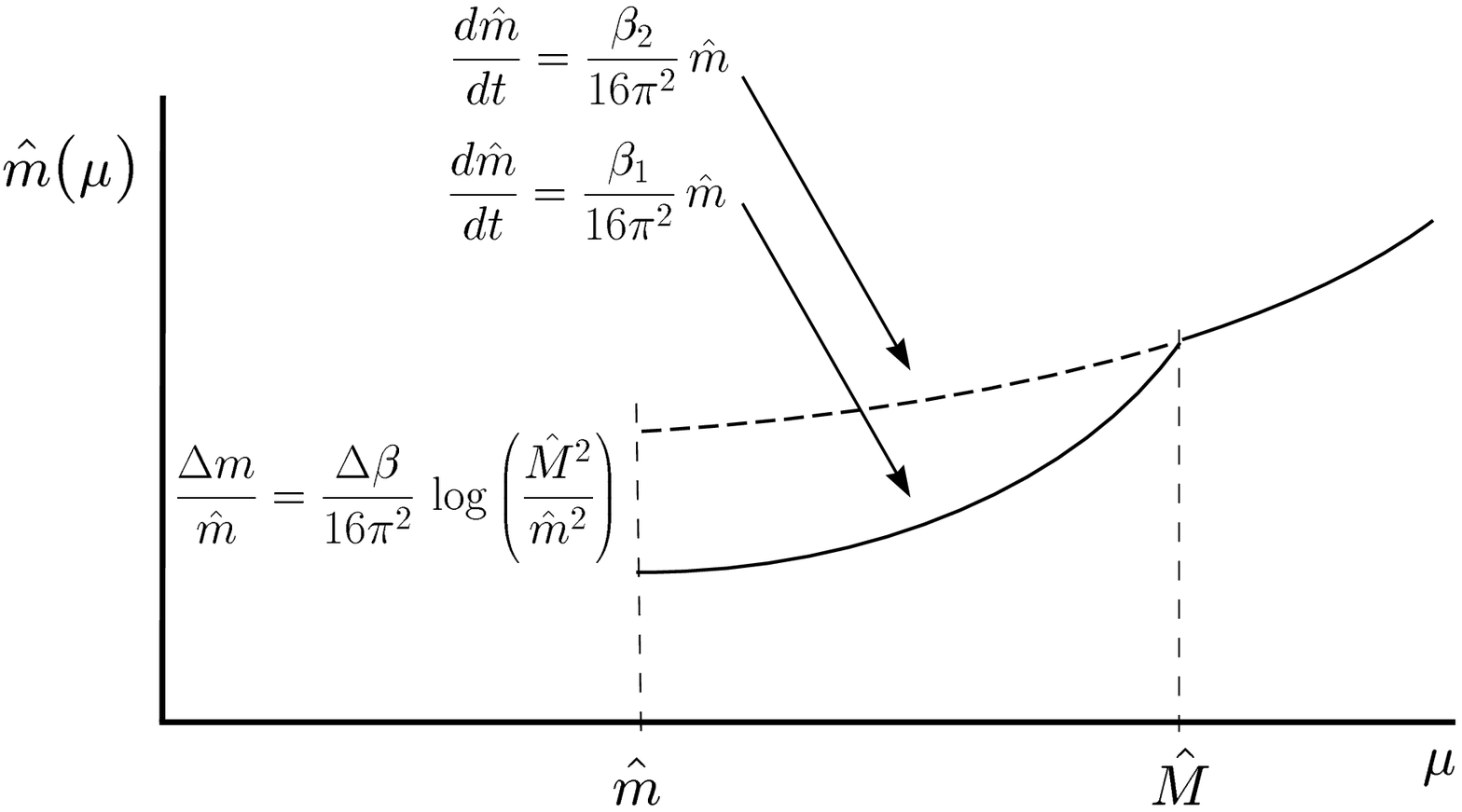}
\fcaption{The \dr mass $\hat m(\mu)$ vs. the \dr scale $\mu = e^t$.}
\end{figure}

At the weak scale, we specify the \dr gauge couplings $\hat g_1(M_Z)$
and $\hat g_2(M_Z)$, and the \dr vev $\hat v$, which we extract
from the Fermi constant, $G_F$, the $Z$-boson mass, $M_Z$, and the
electromagnetic coupling $\alpha_{\rm EM}$.  We also specify the top
and tau \dr Yukawa couplings $\hat\lambda_t(M_Z)$ and $\hat\lambda_\tau
(M_Z)$, which we compute from $\tan\beta \equiv \tan\hat\beta =
\hat v_2/\hat v_1$ and the top and tau pole masses, $m_t$ and $m_\tau$.

We define $M_{\rm GUT}$ to be the scale where $\hat g_1(M_{\rm GUT})
= \hat g_2(M_{\rm GUT})$.  At $M_{\rm GUT}$ we
specify the \dr supersymmetric parameters $M_{1/2} \equiv \hat M_{1/2}
(M_{\rm GUT})$, $M_0 \equiv \hat M_0(M_{\rm GUT})$, and $A_0 \equiv
\hat A_0(M_{\rm GUT})$.  We also define the \dr strong coupling $\hat
g_3(M_{\rm GUT})$ by the unification condition
\begin{equation}
\hat g_3(M_{\rm GUT})\ =\ \hat g_1(M_{\rm GUT})
(1 + \epsilon_g)\ ,
\label{eq:epsg}
\end{equation}
where $\epsilon_g$ parametrizes the gauge-coupling threshold correction
at the unification scale.  In a similar way, we impose bottom-tau
unification and define the bottom-quark \dr Yukawa coupling to be
\begin{equation}
\hat\lambda_b(M_{\rm GUT})\ =\ \hat\lambda_\tau(M_{\rm GUT})
(1+\epsilon_b)\ ,
\label{eq:epsb}
\end{equation}
where $\epsilon_b$ parametrizes the unification-scale Yukawa coupling
threshold correction.

We solve the supersymmetric renormalization group equations
self-consistently, subject to the boundary conditions at $M_Z$
and $M_{\rm GUT}$.  In this way we find a prediction for the \dr
couplings $\hat g_3(M_Z)$ and $\hat\lambda_b(M_Z)$.
We then apply the weak-scale threshold corrections to determine
the \ms strong coupling, $\alpha_s(M_Z)$, the bottom-quark
pole mass, $m_b$, and the full supersymmetric
particle spectrum.  Our results allow us to correlate the
unification-scale thresholds $\epsilon_g$ and $\epsilon_b$ with
$m_b$, $\alpha_s(M_Z)$ and the superparticle spectrum.

\section{Threshold Corrections}
\vspace{-0.7cm}
\subsection{The Match-and-Run Procedure}
\vspace{-0.25cm}

To study unification-scale physics, it is important to use the
correct weak-scale threshold corrections.  Typically, these corrections
are computed using the so-called match-and-run technique, which is based
on the successive decoupling of particles at the scale of their masses.
As an example of this technique, let us consider a running \dr mass $\hat
m$, whose  full $\beta$-function depends on all the particles in the
supersymmetric standard model.  For large renormalization scales $\mu$,
$\hat m$ evolves according to its complete supersymmetric RGE.

Let us now run $\hat m$ towards the weak scale, $M_Z$.  Along the
way, we eventually encounter the scale of the squark masses.  According to
the match-and-run procedure, we must stop the evolution and construct a new
effective theory in which the squarks are integrated out.  We must then
continue
the evolution, using the new $\beta$-function, without the squark contribution,
subject to the matching condition $\hat m(m_{sq}^-) = \hat m(m_{sq}^+)$.
We repeat this procedure at each new threshold, finally stopping at the scale
$\mu=\hat m(\mu)$.  The quantity $\hat m(\hat m)$ is the match-and-run
approximation to the physical pole mass.

To test this procedure, let us compute the threshold correction to
a particle of mass $\hat m$ from a particle of mass $M$, with $M > m$.
According to the match-and-run procedure, the decoupling of the heavy
particle gives the correction
\begin{equation}
{\Delta m\over \hat m} \ =\ {\Delta\beta\over16\pi^2}
\ \log\left({\hat M^2\over \hat m^2}\right)\ ,
\end{equation}
as shown in Figure 1.  In this expression, $\Delta\beta$ is the
difference of the $\beta$-functions, before and after decoupling.

\begin{figure}[t]
\epsfysize=1.5in
\hspace*{1.3in}\vspace*{.2in}
\epsffile{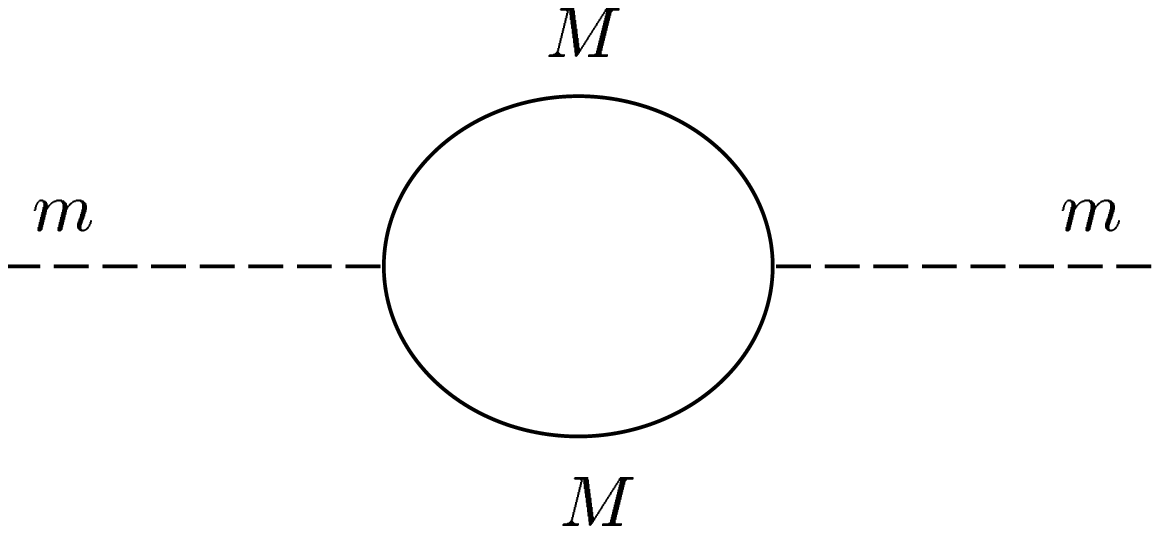}
\fcaption{The one-loop diagram that yields the threshold correction
for a particle of mass $m$ from a particle of mass $M$.}
\end{figure}

The exact one-loop result can be found by computing the diagram of
Figure 2.  It is given by
\begin{equation}
{\Delta m\over \hat m}
\ =\ {\Delta\beta\over16\pi^2}\ \int_0^1\ dx
\ \log\left({\left| \,M^2  - x(1-x)\,m^2\,\right|\over\mu^2}\right)\ .
\label{one loop}
\end{equation}
For $\mu=m$, this reduces to
\begin{equation}
{\Delta m\over \hat m}\ = \ {\Delta\beta\over16\pi^2}\ \biggl[\log\left(
{M^2\over m^2}\right)\ +\  {\rm finite}\ \biggr] \ ,
\end{equation}
where the finite contribution does not contain any logarithms.

These results indicate that the match-and-run procedure gives a good
approximation to the pole mass when $\hat M \gg \hat m$.   In this case the
large logarithm, proportional to $\log(\hat M^2/\hat m^2)$,
dominates the threshold correction.

When $\hat m \simeq \hat M$, however, the finite term is typically as
large as the logarithm.  The finite correction is completely
missed by the match-and-run procedure.  In supersymmetric unified models
with universal boundary conditions, it is quite possible for all the
supersymmetric masses to be near $M_Z$, in which case the
match-and-run procedure gives an error of ${\cal O}(1)$.

\subsection{Gauge Coupling Thresholds}
\vspace{-0.25cm}

It is important to use the correct weak-scale threshold corrections because
supersymmetric unification depends sensitively on the weak-scale boundary
conditions.  This can be easily understood from the one-loop renormalization
group equations for the gauge couplings.  Assuming unification, the RGE's imply
\begin{equation}
{\beta_2-\beta_3\o \hat g_1^2(\mu)}\ +
\ {\beta_3-\beta_1\o \hat g_2^2(\mu)}\ +
\ {\beta_1-\beta_2\o \hat g_3^2(\mu)} \ =\ 0\ ,
\end{equation}
at any scale $\mu$, where the $\beta_i$ are the one-loop $\beta$-functions
for the running gauge couplings.  Solving for $\alpha_s(M_Z)$ and varying
the inputs $\hat g_1(M_Z)$ and $\hat g_2(M_Z)$, one finds
\begin{equation}
{\delta\alpha_s\o\alpha_s(M_Z)} \ \simeq
\ -\ 10.6\ {\delta\hat g_1\o\hat g_1(M_Z)} \ +
\ 12.2 \ {\delta\hat g_2\o\hat g_2(M_Z)} \ .
\label{delta_alpha}
\end{equation}
This implies that an error in the determination of $\hat g_1(M_Z)$ or
$\hat g_2(M_Z)$ of 1\% percent can lead to an error in the evaluation
of $\alpha_s(M_Z)$ of more than 10\%.

The fact the RGE's can naturally amplify small errors means that one
must determine the weak-scale threshold corrections to complete
one-loop accuracy.  Therefore, in our analysis,
we compute $\hat g_1(M_Z)$ and $\hat g_2(M_Z)$
directly from the Fermi constant, $G_F = 1.16639 \times 10^{-5}$ GeV$^{-2}$,
the $Z$-boson mass $M_Z = 91.187$ GeV, the electromagnetic coupling
$\alpha_{\rm EM} = 1/137.036$, the top-quark mass $m_t$, and the parameters
that describe the supersymmetric model.  We include logarithmic and finite
contributions, so our results are more accurate than those based on the
match-and-run procedure.  Note that we do not use the
best-fit value of the standard-model weak mixing angle, $\sin^2
\hat\theta_{\rm SM}$.  This is because the standard-model value of
$\sin^2 \hat\theta_{\rm SM}$, extracted from a combined fit to the data,
cannot be used in a supersymmetric analysis when finite corrections
are important.  (See also [2].  We did not include the complete
gauge coupling thresholds in the preliminary results presented at the
conference.)

We start our analysis by computing the electromagnetic coupling $\hat\alpha$
and the supersymmetric weak mixing angle $\hat s^2$ in the \dr renormalization
scheme,
\begin{equation}
\hat\alpha\ =\ {\alpha_{\rm EM}\o1-\Delta\hat\alpha}
\ ,\qquad\qquad\qquad\qquad \hat{s}^2\hat{c}^2\ = \
{\pi \alpha_{\rm EM}\o\sqrt2 G_F M_Z^2(1-\Delta\hat{r})}\ ,
\end{equation}
where $\hat c^2 = 1 - \hat s^2$,
\begin{equation}
\Delta\hat\alpha\ =\ 0.0658 \pm 0.0007
\ -\ {\alpha_{\rm EM}\o2\pi}\Biggl\{
-{7\o4}\log\left(M_W\o M_Z\right)
\ +\ {16\o9}\log\left(m_t\o M_Z\right)
\ +\ {1\o3}\log\left(m_{H^+}\o M_Z\right)
\label{eq:da}
\end{equation}
$$
\ +\ \sum_{i=1}^6{4\o9}\log\left(m_{\tilde u_i}\o M_Z\right)
\ +\ \sum_{i=1}^6{1\o9}\log\left(m_{\tilde d_i}\o M_Z\right)
\ +\ \sum_{i=1}^3{1\o3}\log\left(m_{\tilde e_i}\o M_Z\right)
+ \sum_{i=1}^2{4\o3}\log\left(m_{\chi_i^+}\o M_Z\right)\Biggr\}\ ,
$$
and\cite{Degrassi}
\begin{equation}
\Delta\hat r\ =\ \Delta\hat\alpha\ +\ {\hat \Pi_W(0)\o M_W^2}
\ -\  {\hat \Pi_Z(M_Z)\o M_Z^2}\ +\ \hbox{vertex} \ + \ \hbox{box}
\ .  \label{eq:dr}
\end{equation}
Equation (\ref{eq:da}) includes the light quark contribution extracted from
experimental data,\cite{Jegerlehner} together with the leptonic contribution.
It also contains the logarithms of the $W$-boson, top-quark, charged-Higgs,
squark, slepton, and chargino masses.  In eq.~(\ref{eq:dr}), the $\hat\Pi$
denote the real and transverse parts of the gauge boson self-energies,
evaluated
in the \dr scheme.  Equation (\ref{eq:dr}) also includes the vertex and box
contributions that renormalize the Fermi constant, as well as the leading
higher-order $m_t^4$ and QCD standard-model corrections given in ref.~5.
(For more details about our calculation see [6].)

{}From these results we find the weak gauge couplings $\hat g_1(M_Z)$ and
$\hat g_2(M_Z)$ using the \dr relations
\begin{equation}
\hat g_1(M_Z)\ =\ \sqrt{5\o3}\,{\hat e\o\hat c}\ ,\qquad\quad
\hat g_2(M_Z)\ =\ {\hat e\o\hat s}\ ,
\end{equation}
where $\hat\alpha = \hat e^2/4\pi$.
These couplings serve as the weak-scale boundary conditions for the two-loop
supersymmetric RGE's.\cite{two loop}  Note that they depend on the masses
of the supersymmetric particles, which are determined self-consistently
through the solution to the renormalization group equations.  The \dr
values for $\hat g_1(M_Z)$ and $\hat g_2(M_Z)$ contain the full one-loop
threshold corrections, including the logarithmic and finite contributions.

\subsection{Mass Thresholds}
\vspace{-0.25cm}

Since the exact one-loop self-energies have been computed for all
particles in the minimal supersymmetric standard model,\cite{corrs}
it is possible to include all the logarithmic and finite
corrections in the threshold corrections.  In this section we will
discuss the leading corrections for the top- and bottom-quark masses.

The relation between the top-quark pole mass and the running \dr mass is given
by
\begin{equation}
m_t\ =\ \hat m_t(\mu) \ -\  \hat\Sigma_t(m_t)\ ,
\label{dr-pole}
\end{equation}
where $\hat m_t$ is the \dr mass, and $\hat\Sigma_t$ is the top-quark
self-energy,
defined by $\hat\Sigma_t = \hat\Sigma_1 + m_t\hat\Sigma_\gamma$, and the quark
self-energy is written $\hat\Sigma_1 + \hat\Sigma_\gamma\rlap/p +
\hat\Sigma_{\gamma5}
\rlap/p\gamma_5  + \hat\Sigma_5\gamma_5$.

Equation (\ref{dr-pole}) relates the running mass to the pole mass
at {\it any} scale $\mu$ near $m_t$.  In a unification analysis,
this leads to a computational
simplification because the RGE's can be run down to a single scale $\mu$, and
then all the threshold corrections applied simultaneously.  For the case at
hand,
we take the scale $\mu = M_Z$.

The top quark threshold correction receives its most important contributions
from gluon and gluino/squark loops.  The correction from the gluon loop is
well-known,\footnote{Actually this correction is more commonly seen as
$4\alpha_s/(3\pi)$, which is the result for $\mu=m_t$ in the \ms scheme.}
\begin{equation}
m_t\ =\ \hat m_t(\mu)\ \left\{1 \ +\ {\alpha_s\over3\pi}\left[ 5 + 6 \log
\left({\mu\over m_t}\right)\right]\right\}\ ,
\label{gl}
\end{equation}
where $\hat m_t(\mu)$ is the running \dr mass evaluated at the scale
$\mu$.  For $\mu = m_t$, this correction is about 6\%.  The gluino/squark
contribution is given by
\begin{eqnarray}
\Delta m_t^{\tilde{g}\tilde{q}}\ &=&\  -{\alpha_s\over3\pi}\ \Biggl\{
{\rm Re}\left[B_1(m_t,m_{\tilde{g}},m_{\tilde{t}_1})
+ B_1(m_t,m_{\tilde{g}},m_{\tilde{t}_2}) \right]  \label{gl:sq}\\
&&-\ {2m_{\tilde{g}}\left(A_t+\bar\mu\cot\beta\right)\over
m_{\tilde{t}_1}^2 - m_{\tilde{t}_2}^2}
\,{\rm Re}\left[ B_0(m_t,m_{\tilde{g}},m_{\tilde{t}_1})-
B_0(m_t,m_{\tilde{g}},m_{\tilde{t}_2}) \right]\Biggr\}\ \hat m_t(\mu)\
,\nonumber\\
\nonumber
\end{eqnarray}
where $B_0$ and $B_1$ are the two point functions
\begin{equation}
B_n(p,m_1,m_2)\ =\ -\int_0^1dx\,x^n\,\log\left(
{(1-x)\,m_1^2 + x\,m_2^2 - x(1-x)\,p^2\over
\mu^2}\right)\ ,
\end{equation}
and $\bar\mu$ is the Higgs mass parameter.
This expression contains a finite plus a logarithmic piece for $m_{\tilde{g}}
> m_t$ and/or $m_{\tilde{t}}>m_t$.  The logarithmic contribution, which can
be larger than the gluon contribution (\ref{gl}), is given
correctly by the match-and-run procedure.  The finite piece is of order
1\%.

The bottom-quark mass also receives a significant threshold correction,
similar to (\ref{gl}), from the diagram with a gluon loop.  For $\mu = M_Z$,
the large logarithm $\log(m_b/M_Z)$ must be resummed to determine the
pole mass.  For the bottom quark, the correction from the gluino/squark
diagram is also important; the second line in (\ref{gl:sq}) becomes
\begin{equation}
{\Delta m_b\over m_b}\ \simeq\ -{2\alpha_s\over 3\pi} {\bar\mu
m_{\tilde{g}}\over m_{\tilde{b}}^2}\ \tan\beta\
\label{bottom:finite}
\end{equation}
for large $\tan\beta$.
This correction,\cite{mbtanbeta} which is completely missed in the
match-and-run
procedure, can be as large as 50\%.  In the following, we choose $\bar\mu>0$
to minimize the bottom-quark mass.

\subsection{Yukawa Coupling Thresholds}
\vspace{-0.25cm}

To discuss supersymmetric unification, and in particular, bottom-tau Yukawa
coupling unification, one needs to find the running \dr Yukawa couplings.
In this section we will show how to use the mass threshold corrections to
find the running \dr Yukawa couplings.  The \dr Yukawa couplings contain
all the weak-scale threshold corrections.

\begin{figure}[t]
\epsfysize=3.5in
\epsffile[80 440 500 715]{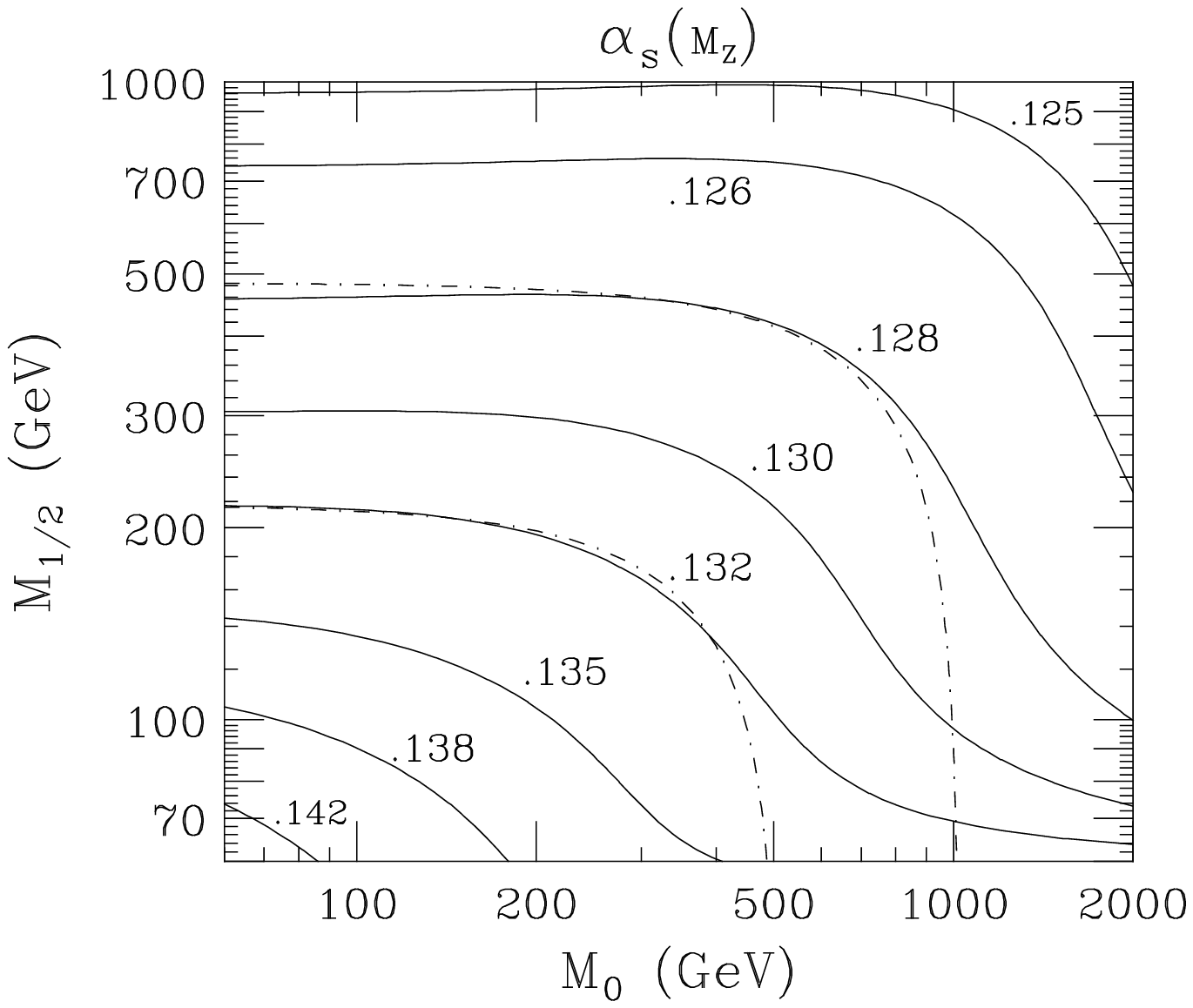}
\vspace{.25in}
\fcaption{Contours of $\alpha_s(M_Z)$ and one-loop light-squark masses
in the $M_0$, $M_{1/2}$ plane, with $m_t=170$ GeV, $\tan\beta=2$ and $A_0=0$.
The dot-dashed contours indicate squark masses of 500 and 1000 GeV.
The squark masses obey the heuristic relation
$m_{\tilde q}^2 \ \simeq\ M_0^2 + 4.5\, M_{1/2}^2$. }
\end{figure}

Let us illustrate our procedure for the case of the top quark.  As above,
we find the running \dr mass from the physical pole mass using (\ref{dr-pole}).
We then relate the \dr mass to the \dr vev and Yukawa coupling by the simple
relation
\begin{equation}
\hat m_t(\mu) \ =\ {1\over\sqrt{2}}\hat\lambda_t(\mu) \hat v_2(\mu)\ .
\end{equation}
The \dr vev is determined from the gauge couplings and the
$Z$-boson mass,
\begin{equation}
M_Z^2\ =\ {1\over4}\,(\hat g(\mu)^2 + \hat g'(\mu)^2)\,\hat v^2
\ -\  \hat \Pi_Z(M_Z)\ .
\end{equation}
Combining these expressions, we find the threshold-corrected
Yukawa coupling,
\begin{equation}
\hat\lambda_t(\mu)\ =\ {\hat g_2(M_Z) \over \sqrt{2}\, \hat c}\,{m_t \over M_Z}
\ \left[ \,1 + {\hat \Sigma_t(m_t)\over m_t} - {1\over2}\,
{\hat \Pi_Z(M_Z)\over M^2_Z}\,\right]\ ,
\label{eq:lambda}
\end{equation}
where $m_t$ is the top-quark pole mass.
Equation (\ref{eq:lambda}) converts the measured top-quark mass
into a threshold-corrected Yukawa coupling at the scale $\mu$.

Following this procedure, we can find all the threshold-corrected Yukawa
couplings in terms of two-point functions.  The great
advantage of this approach is that the final formulae are much simpler
than those obtained using three-point diagrams.

\section{Numerical Analysis}
\vspace{-0.7cm}
\subsection{Gauge Coupling Unification}
\vspace{-0.25cm}

Now that we have the full set of one-loop threshold corrections, we can
carry out a consistent numerical evaluation of the two-loop supersymmetric
RGE's.  In what follows we will study correlations between the low-energy
supersymmetric spectrum and the physics of the unification scale.

As a point of reference, we show in Figure 3 our results for $\alpha_s(M_Z)$
in the $M_0$, $M_{1/2}$ plane, with no unification-scale threshold
corrections, for $m_t=170$ GeV, $\tan\beta=2$, and $A_0=0$.  Comparing
with the value from the particle data group,\cite{PDG} $\alpha_s(M_Z) =
0.117 \pm 0.005$, we see that $\alpha_s(M_Z)$ is rather large.

The value of $\alpha_s(M_Z)$ is correlated with the low-energy
supersymmetric spectrum through its dependence on $M_0$ and $M_{1/2}$.
In Figure 3 we also show representative values of the light-squark masses
in the $M_0$, $M_{1/2}$ plane.
If, for naturalness, we require the squark masses to be below 1 TeV, we
find the lower bound $\alpha_s(M_Z) > 0.127$ from Figure 3.

The values of $\alpha_s(M_Z)$ quoted here are larger than in many previous
analyses for two reasons.  First, during the past few years, the best-fit
value for the top-quark mass has been increasing.  This implies a decreasing
central value for the standard-model weak mixing angle, which then leads to
an increase in $\alpha_s(M_Z)$.
Second, the finite corrections decrease $\hat s^2$, which leads to a further
increase in $\alpha_s(M_Z)$.  The finite corrections are important in the
region
$M_{1/2}\ \roughly{<}\ 200$ GeV where $\alpha_s(M_Z)$ is appreciably larger
than in the leading logarithmic approximation, as shown in Figure 4.

\begin{figure}[tb]
\epsfysize=2.8in
\epsffile[100 510 680 710]{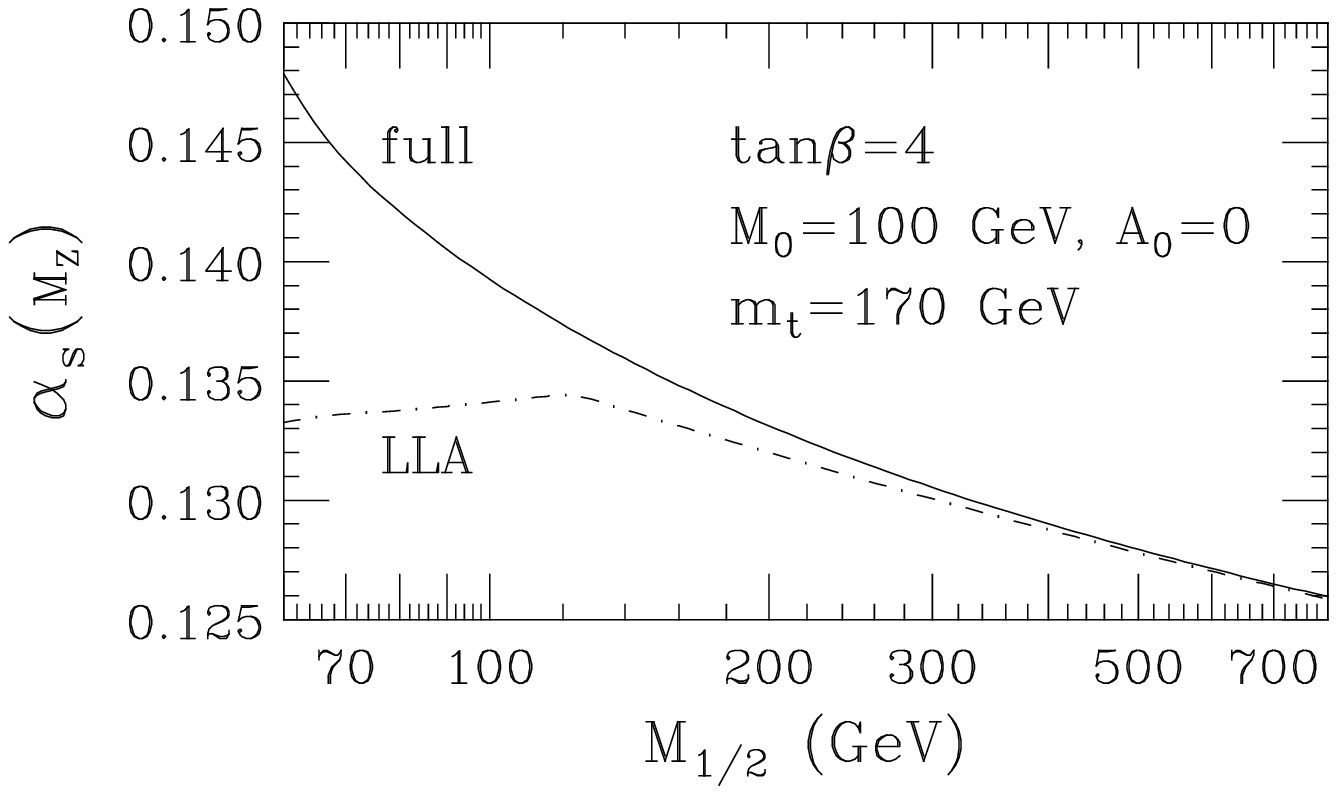}
\vspace{.25in}
\fcaption{The \ms coupling $\alpha_s(M_Z)$ vs. $M_{1/2}$.
The curve labeled LLA shows the result if we include
only the logarithms of the supersymmetric masses (the leading logarithm
approximation), while the solid line corresponds to the full result
including all finite corrections.}
\end{figure}

Of course, the value of $\alpha_s(M_Z)$ can be reduced by a unification-scale
threshold correction with $\epsilon_g  < 0$.  This is illustrated in Figure 5,
where we show the upper and lower bounds on $\epsilon_g$ necessary to obtain
$\alpha_s(M_Z) =0.117 \pm 0.01$.  We find that $\epsilon_g$ of just
$-2$\% is sufficient for a wide range of supersymmetric masses.  For squark
masses of order 1 TeV, no unification thresholds are required.

\begin{figure}[tb]
%\epsfysize=1.75in
%\epsffile[30 537 800 702]{Warsaw5.ps}
\epsfysize=2in
\epsffile[50 540 850 715]{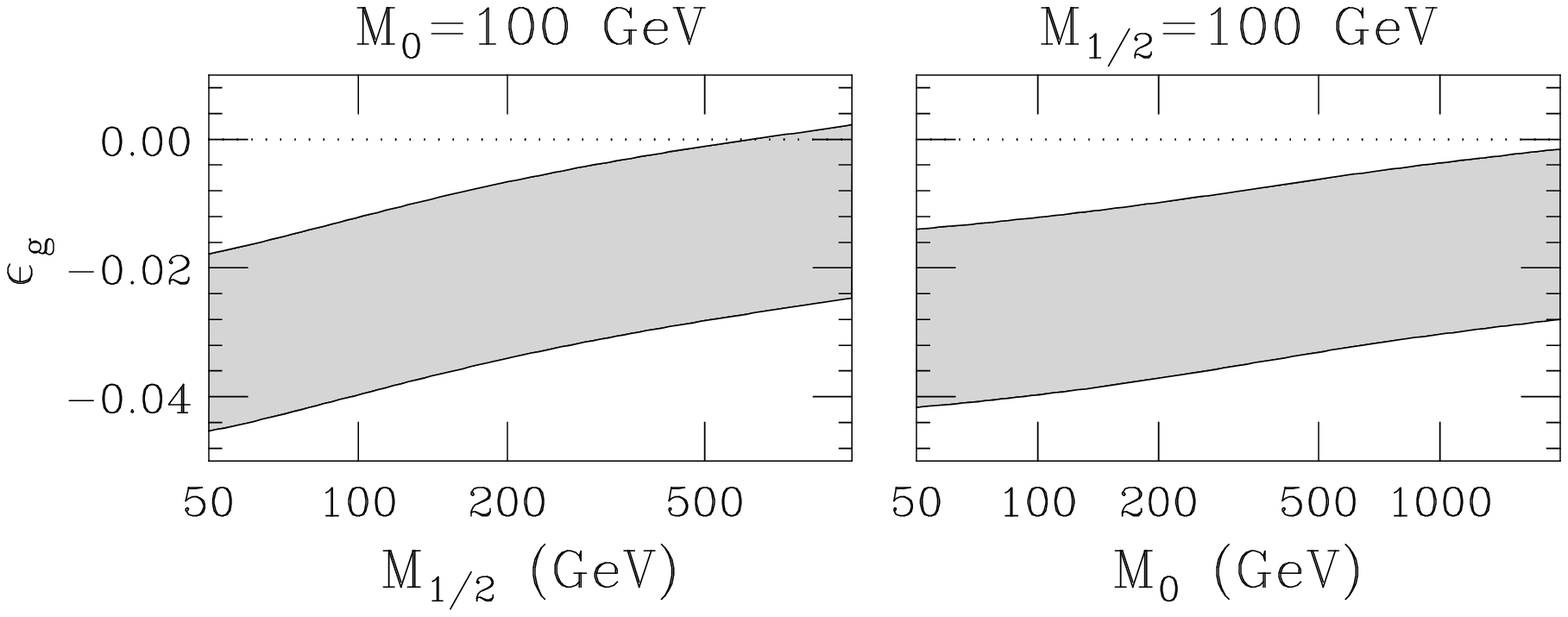}
\fcaption{The shaded regions indicate the allowed
values of the gauge coupling threshold correction, $\epsilon_g$,
necessary to obtain $\alpha_s(M_Z) = 0.117\pm 0.01$, with
$m_t = 170$ GeV, $\tan\beta = 2$ and $A = 0$.}
\end{figure}

\subsection{Yukawa Coupling Unification}
\vspace{-0.25cm}

In typical SU(5) models, the bottom and tau Yukawa couplings unify at the
scale $M_{\rm GUT}$.  This leads to a prediction for the bottom-quark pole
mass, $m_b$, as a function of $m_t$ and $\tan\beta$.  For $m_t \simeq 170$
GeV and arbitrary $\tan\beta$, one typically finds a bottom-quark mass above
the range experiment,\cite{PDG} which we take to be $4.7 < m_b < 5.2$ GeV.
There are two regions of $\tan\beta$ where $m_b$ is smaller and
successful bottom-tau unification is easier to achieve:
$\tan\beta\ \roughly{<}\ 2$ and $\tan\beta \ \roughly{>}\ 40$.

In the region $\tan\beta\ \roughly{<}\ 2$, the top-quark Yukawa
coupling is large, approaching the perturbative limit.  The large coupling
drives
electroweak symmetry breaking and significantly
affects the RGE for the bottom-quark Yukawa.
It pushes $\hat\lambda_b(M_Z)$ down and decreases the bottom-quark mass.
In the region of large $\tan\beta$, both the top- and bottom-quark Yukawa
couplings
are large.  They both make significant contributions to the RGE's.  Indeed, for
large
$\tan\beta$, the large bottom-quark Yukawa actually drives {\it down} the value
of the bottom-quark mass.  Therefore we have the possibility of successful
bottom-tau unification for small and large $\tan\beta$.

In Figure 6 we show $m_b$ and $\alpha_s(M_Z)$ versus $m_t$, for small
$\tan\beta$
and various values of $M_0$, and $M_{1/2}$, with no unification-scale
thresholds.
The figure corresponds to $\hat\lambda_t(M_{\rm GUT}) = 3$.  Smaller
values of $\hat\lambda_t(M_{\rm GUT})$ give rise to larger values of $m_b$, so
the curves can be interpreted as lower limits on the bottom-quark mass
(in the small $\tan\beta$ region).  We see that $m_b$ is large unless
the squark masses are of order 1 TeV.

\begin{figure}[tb]
\epsfysize=3in
\epsffile[5 395 680 690]{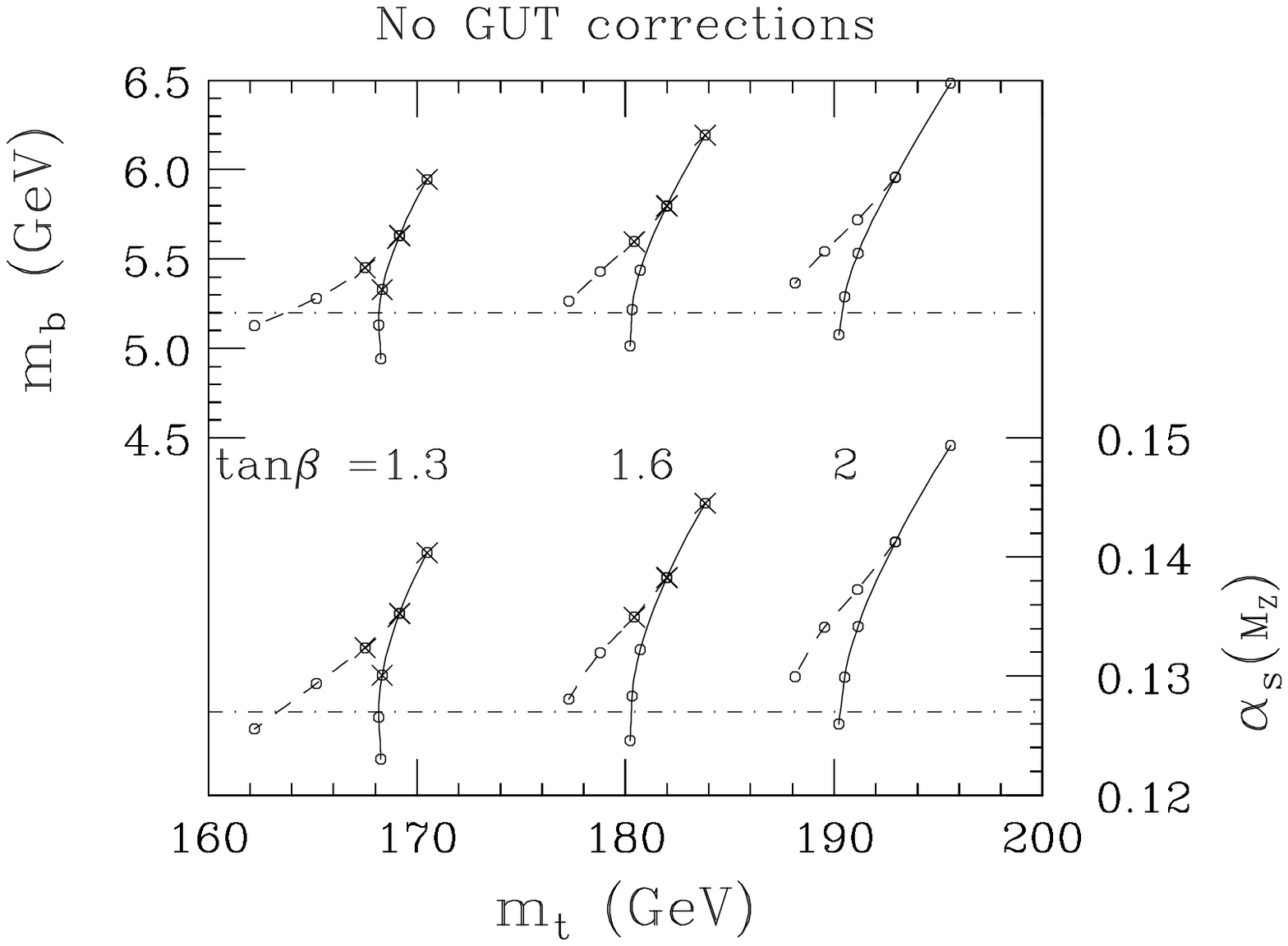}
\vspace{.2in}
\fcaption{The bottom-quark pole mass and $\alpha_s(M_Z)$
vs. $m_t$, with no unification-scale thresholds,
for various values of $\tan\beta$, for $A_0=0$ and
$\hat\lambda_t(M_{\rm GUT})=3$. The right (solid) leg in
each pair of lines corresponds
to $M_{1/2}$ varying from 60 to 1000 GeV, with $M_0$ fixed at 60 GeV.
The left (dashed) leg corresponds to $M_0$ varying from
60 to 1000 GeV, with $M_{1/2}=100$ GeV. On the solid lines the circles
mark, from top to bottom, $M_{1/2}=60$, 100, 200, 400, and 1000 GeV,
and on the dashed lines the circles mark $M_0=60$, 200, 400, and
1000 GeV. The lowest point on each left leg and the
second-to-lowest point on each right leg corresponds to
$m_{\tilde q}\simeq1$ TeV. The $\times$'s mark points
with one-loop Higgs mass $m_h<60$ GeV.}
\end{figure}

\begin{figure}[tb]
\epsfysize=3in
\epsffile[0 408 700 708]{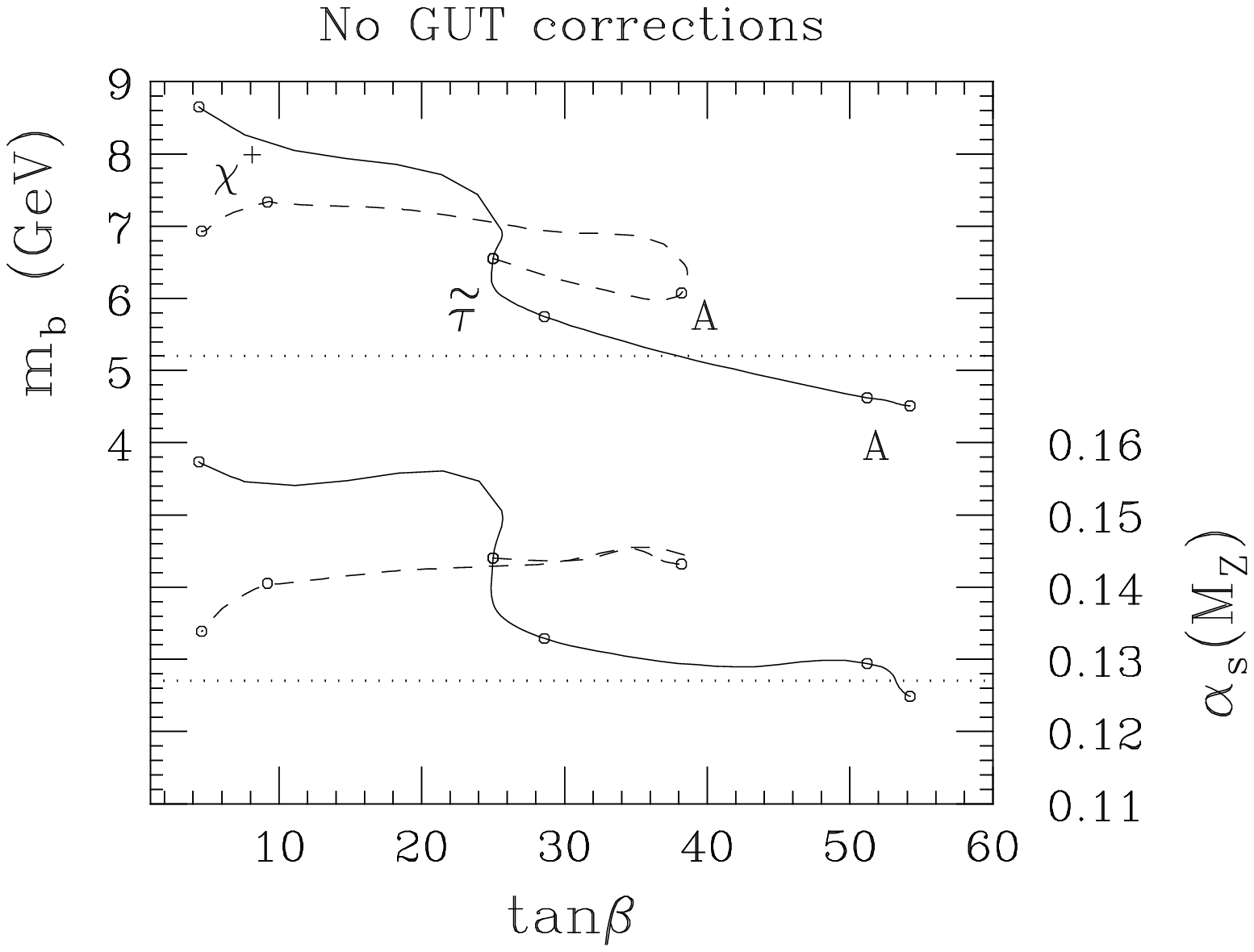}
\fcaption{The bottom-quark pole mass and $\alpha_s(M_Z)$
vs. the maximum $\tan\beta$, with no unification-scale
thresholds, for $m_t = 170$ GeV and $A_0=0$. The solid line
corresponds to $M_{1/2}$ varying from 60 to 1000 GeV, with $M_0$
fixed at 60 GeV.  The dashed line corresponds to $M_0$ varying from
60 to 1000 GeV, with $M_{1/2}=100$ GeV. On the solid lines the circles
mark, from top to bottom, $M_{1/2}=60$, 100, 200, 400, and 1000 GeV,
and on the dashed lines the circles mark $M_0=60$, 200, 400, and
1000 GeV. The maximum value of $\tan\beta$ is determined by the
experimental limits on the masses of the $A$, $\tilde\tau$ and the
$\chi^+_1$, as indicated.}
\end{figure}

In Figure 7 we show $m_b$ and $\alpha_s(M_Z)$ versus $\tan\beta$, in
the large $\tan\beta$ region, for $m_t = 170$, and various values of $M_0$, and
$M_{1/2}$, with no unification-scale thresholds.  For each point, we
choose the maximum value of $\tan\beta$ subject to the physical requirements
$m_A > 22$ GeV, $m_{\tilde\tau} > 45$ GeV and $m_{\chi_1^+} > 47$ GeV,
where $A$ is the CP-odd Higgs boson, $\tilde\tau$ is the tau slepton,
and $\chi^+_1$ is the lightest chargino.  From the figure we see that
the smallest values of $m_b$ occur for squark masses near 1 TeV.

As with $\alpha_s(M_Z)$, the picture is changed by unification-scale threshold
corrections.  To understand their effects, note the striking similarity between
the $m_b$ and $\alpha_s(M_Z)$ curves in Figures 6 and 7.  This tells us that
the value
of $m_b$ is tightly correlated with the value of $\alpha_s(M_Z)$.  It leads us
to
expect that the gauge threshold correction will have an important effect on
$m_b$.

This expectation is confirmed in Figure 8, where we show the band of
unification-scale
thresholds $\epsilon_b$ that are necessary to bring $m_b$ into the preferred
range.
For the figure, we first choose $\epsilon_g$ to fix $\alpha_s(M_Z)$ at its
central
value, $\alpha_s(M_Z) = 0.117$.  We then vary $\epsilon_b$ to obtain
$4.7 < m_b < 5.2$.  The two bands of $\epsilon_b$ correspond to the regions
of small and large $\tan\beta$.  In the first region, we set $\tan\beta = 2.5$,
and in the second, $\tan\beta=30$.  The figure indicates that Yukawa threshold
corrections of less than 10\% are sufficient to ensure successful unification,
provided the gauge thresholds are such that $\alpha_s(M_Z)$ agrees with
experiment.

\begin{figure}[tb]
\epsfysize=3.2in
\epsffile[30 425 650 735]{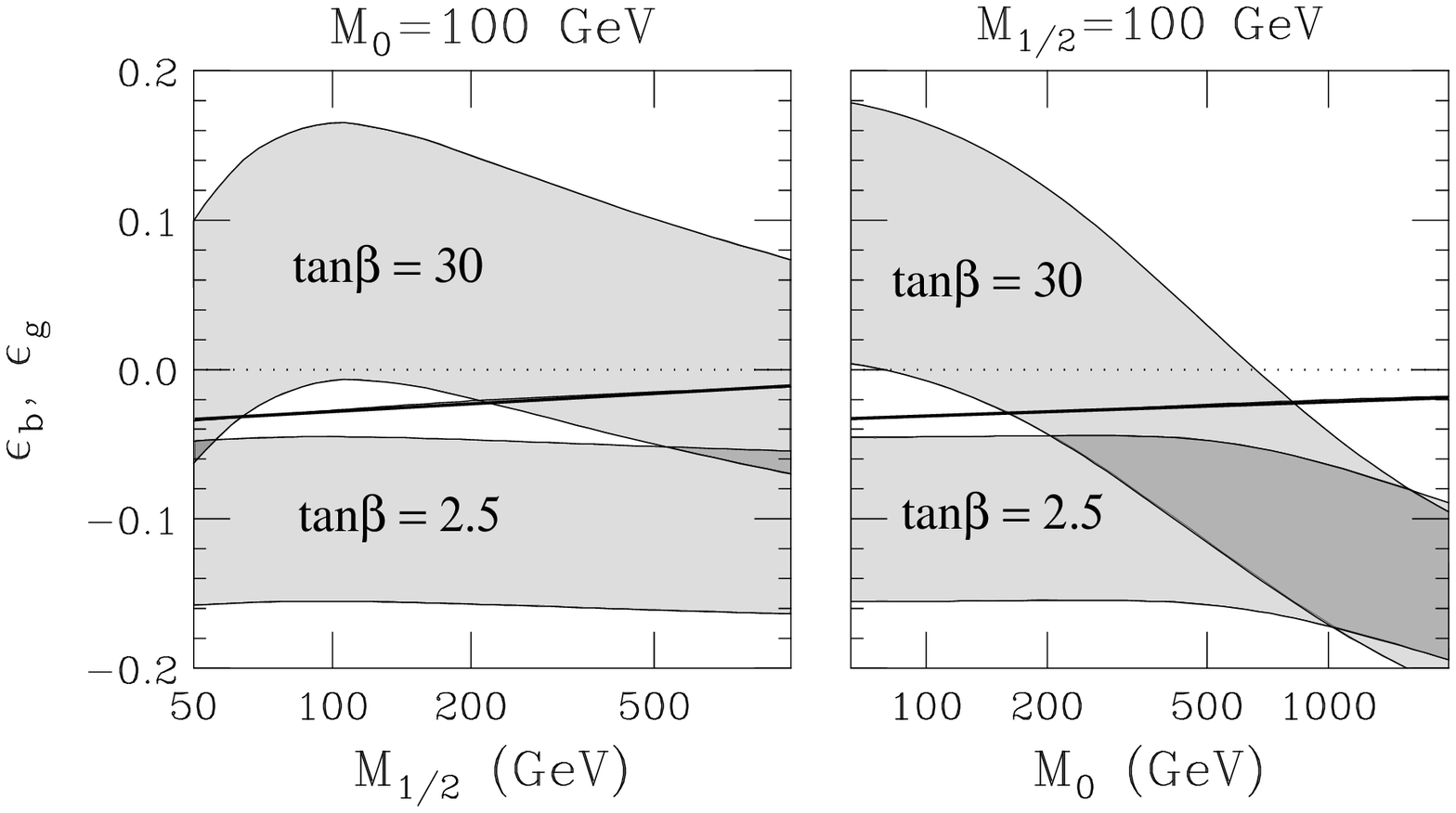}
\fcaption{Values of the bottom-quark Yukawa coupling
threshold correction, $\epsilon_b$, necessary to obtain $4.7 < m_b < 5.2$,
assuming $\epsilon_g$ is adjusted to give $\alpha_s(M_Z) = 0.117$.
The solid line marks $\epsilon_g$, while the two bands indicate
$\epsilon_b$ for different values of $\tan\beta$, for $A_0 = 0$ and
$m_t = 180$ GeV.}
\end{figure}

\section{Conclusions}
\vspace{-0.25cm}

In this talk we discussed the complete one-loop weak-scale threshold
corrections in the minimal supersymmetric standard model.  We illustrated
how they affect gauge and Yukawa unification, with and without
unification-scale
threshold corrections.  In the absence of such corrections, we find that
$\alpha_s(M_Z)$ and $m_b$ are large unless the squark masses are
of order one TeV.  With small threshold corrections, we find acceptable
values for $\alpha_s(M_Z)$ and $m_b$ for heavy or light supersymmetric
spectra.

We would like to thank R.~Zhang for collaboration during the
early stages of this work, and
S.~Pokorski for helpful discussions.  J.B. would like to thank Z.~P\l uciennik
for extensive discussions  during the conference.  This work was
supported by the U.S. National Science Foundation under grant NSF-PHY-9404057.

\end{document}